\documentclass[twocolumn,showpacs,amsmath,amssymb]{revtex4}

\usepackage{graphicx}
\usepackage{dcolumn}
\usepackage{bm}

\begin{document}
\title{Physics of collisionless phase mixing}
\author{D. Tsiklauri}
\author{T. Haruki}

\affiliation{Joule Physics Laboratory,
Institute for Materials Research, University of Salford, Manchester, M5 4WT,
United Kingdom}
\date{\today}

\begin{abstract}
Previous studies of phase mixing of ion cyclotron (IC), Alfv\'enic, waves in the collisionless
regime have established the generation of parallel electric field and hence
acceleration of electrons in the regions of transverse density inhomogeneity.
However, outstanding issues were left open. 
Here we use 2.5D, relativistic, fully electromagnetic PIC (Particle-In-Cell) 
code and an analytic MHD (Magnetohydrodynamic) formulation, to establish the following points:
(i) Using the generalised Ohm's law we find that the parallel electric field is supported mostly by
the electron pressure tensor, with a smaller contribution from the electron inertia term.
(ii) The generated parallel electric field and the fraction of accelerated electrons are 
independent of the IC wave
 frequency remaining at a level of six orders of magnitude larger 
 than the Dreicer value and approximately
20 \% respectively. The generated parallel electric field and the fraction of accelerated electrons 
increase with the increase of IC wave  amplitude. The generated parallel electric field seems to be
independent of  plasma beta, while the fraction of accelerated electrons strongly increases 
with the decrease of plasma beta (for plasma beta of 0.0001 the fraction of accelerated electrons
can be as large as 47 \%).
(iii) In the collisionless regime 
IC wave dissipation length (that is defined as the distance over which the wave damps) 
variation with the driving frequency shows a deviation from the
analytical MHD result, which we attribute to 
a possible frequency  dependence  of the effective resistivity.
(iv) Effective anomalous resistivity, inferred from our numerical simulations, is at least four orders of
magnitude larger than the classical Spitzer value.
\end{abstract}	


\maketitle

\section{Introduction}

Phase mixing is a mechanism of enhanced dissipation of 
Alfv\'en waves due to inhomogeneity of Alfv\'en speed in a
direction transverse to a local magnetic field.
This mechanism originally was studied in the
fusion and laboratory plasma context by a number of 
authors \cite{1972PhFl...15.1673U,1973ZPhy..261..203T,1973ZPhy..261..217G,1974PhRvL..32..454H,
1974PhFl...17.1399C,1975JPlPh..13...87T} and subsequently applied to
the solar corona \cite{1983A&A...117..220H}. Most of the 
large amount of work done in the field of phase mixing 
was in the resistive MHD (Magnetohydrodynamic) regime. Recently, a few works looked at the
same mechanism in the collisionless regime in the context of Earth 
magnetosphere \cite{1999JGR...10422649G, 2004AnGeo..22.2081G} and solar corona
\cite{2005A&A...435.1105T,2005NJPh....7...79T}. The main findings of these
works include the generation of electric field that is parallel to the ambient 
magnetic field in the regions of transverse density inhomogeneity,
as well as associated electron acceleration. It should be mentioned that
these studies considered circularly polarised ion cyclotron (IC) waves
which in the low frequency regime become Alfv\'en waves. 
We use terms Alfv\'en or IC interchangeably, but reader should bear in mind
we always refer to {\it waves with frequencies $< \omega_{ci}$} (with $\omega_{ci}$ being ion cyclotron frequency).
The exact mechanism of generation of the parallel electric field has stimulated
a debate \cite{2006A&A...449..449M,2007NJPh....9..262T}, and even MHD regime option was explored
\cite{2006A&A...455.1073T}.
Continuing this investigation, here we apply technique used in the
collisionless reconnection \cite{pritchett01,th08}.
Namely, in section 3.1 we use generalised Ohm's law to find out
which term generates the parallel electric field.

Solar flare observations \cite{2005SSRv..121..141F} trigger one's 
interest in how effectively plasma particles are accelerated.
Hence, in section 3.2 we look into how 
the generated parallel electric field and the fraction of accelerated electrons
depend on model parameters such as IC wave frequency, amplitude, and plasma beta.

Ref. \cite{1983A&A...117..220H} provides a simple analytical expression
how Alfv\'en wave amplitude should decay in space due to the phase mixing.
Despite the fact that their formula is derived in the resistive MHD
regime, we still apply it to our collisionless, kinetic simulation
and see what does the comparison yield (Section 3.3). 
This is done in the light of previous results of \citet{2005A&A...435.1105T}
who established that in the collisionless, kinetic regime
Alfv\'en wave amplitude in the density gradient regions decays with
distance (from where it is driven) according to collisional MHD formula of 
\citet{1983A&A...117..220H}. Here we stretch the 
MHD-kinetic analogy further to test $\omega_d^2$ dependence under the exponent.

In Sect 3.4 we estimate the effective "resistivity" (again the spirit of MHD-kinetic analogy).
The quotation marks are needed to signify that PIC (Particle-In-Cell)
simulation code is collisionless and hence no resistive effects
exist as such. However, scattering of particles by magnetic
fields plays effective role of collisions.

\section{Simulation Model}

In our numerical simulations we use 2.5D, relativistic, fully electromagnetic PIC code. 
The initial conditions, basic parameters and boundary conditions are exactly the same as 
in the previous work by \citet{2005A&A...435.1105T}.
In particular, 
the uniform magnetic field is in $x$-direction, the transverse density inhomogeneity is across $y$-direction.
$z$ is the spatially ignorable coordinate, i.e. $\partial / \partial z = 0$. However, we retain
all three components of velocity, $v_x$, $v_y$ and $v_z$. 
The system size without ghost cells 
in two dimensions is $L_x = 5000 \Delta$ and $L_y = 200 \Delta$ where $\Delta (= 1)$ is 
the simulation grid size, corresponding 
to the electron Debye length, $\lambda_D = v_{te} / \omega_{pe} = 1 \Delta$ ($v_{te}$ is 
electron thermal velocity and $\omega_{pe}$ is electron plasma frequency).
The total particle number is $2.39 \times 10^8$ electron-ion pairs.
The ion to electron mass ratio is $m_i / m_e = 16$ due to the limitation of 
computer resources and speed.
\citet{2005A&A...435.1105T} use the fixed driving wave 
amplitude  $\delta B / B_0 = 0.05$, and plasma $\beta = 0.02$.
 The dimensionless 
ion and electron density inhomogeneity is described by
$$
 {n_i(y)}=
{n_e(y)}=1+3 \exp\left[-\left(\frac{y-100\Delta}{50 \Delta}\right)^6\right]
\equiv F(y).
$$
These are normalised to some background constant value ($n_0$).
Here all plasma parameters are quoted at the boundary, 
away from the density inhomogeneity region.
In the central region (across $y$-coordinate), the density is
smoothly enhanced by a factor of 4, and there are the 
strongest density gradients having 
a width of about ${51 \Delta}$ around the 
locations $y=51.5 \Delta$ and $y=148.5 \Delta$.
Below, in Eqs.(4) and (5) we shall be using $y=51.5-51/2=26\Delta$
and $y=51.5+51/2=77 \Delta$ as the boundaries of one of the
transverse density gradients.
The background temperature of ions and electrons, 
and their thermal velocities
are varied accordingly
$$
{T_i(y)}/{T_0}=
{T_e(y)}/{T_0}=F(y)^{-1},
$$
$$
 {v_{th,i}}/{v_{i0}}=
{v_{th,e}}/{v_{e0}}=F(y)^{-1/2},
$$
such that the thermal pressure remains constant. Since the background magnetic field
along the $x$-coordinate  is also constant, the total pressure remains constant too.
Then we impose a current of the following form
$$
{\partial_t E_y}=-J_0\sin(\omega_d t)\left(1-\exp\left[-(t/t_0)^2\right]\right),
$$
$$
{\partial_t E_z}=-J_0\cos(\omega_d t)\left(1-\exp\left[-(t/t_0)^2\right]\right).
$$
In Ref. \cite{2005A&A...435.1105T} the driving frequency was fixed at $\omega_d=0.3\omega_{ci}$.
Here,  we also use driving frequencies that satisfy 
$\omega_d < \omega_{ci}$ so that no significant ion-cyclotron resonant 
damping takes place. $\partial_t$ denotes the time derivative.
$t_0$ is the onset time of the driver, which was fixed at $50 /\omega_{pe}$
i.e.  $3.125 / \omega_{ci}$. This means that the driver onset time is about 3 ion-cyclotron
periods. Imposing such a current on the system results in the generation of
left circularly polarised IC (Alfv\'enic) wave, which is driven at the left 
boundary of simulation box and has spatial driver width of $1 \Delta$.
The wave propagates along $x$-coordinate and generates the parallel
electric field in the density gradient regions (for more details see \citet{2005A&A...435.1105T}).
The parameters used are commensurate to what is seen in solar
corona by e.g. Doppler broadening of emission lines.
The observed values of the Alfv\'en waves at heights of 
$R = 1.04 R_{\rm sun}= 28$ Mm are about 
50 km s$^{-1}$ (see e.g. \cite{moran01}), 
which for a typical Alfv\'en speed of 
1000 km s$^{-1}$ makes $\delta B / B_0$ equal to 0.05.

As one of our goals is to investigate the parameter space of the
problem, the following range of physical parameters was used:
 we vary the frequency, amplitude of IC wave or the plasma beta.
The driving wave frequency $\omega_d / \omega_{ci}$ ranged from $0.1$ to $0.5$ with a step of $0.1$.
The wave amplitude $\delta B / B_0$ was also set at $0.01, 0.05, 0.10, 0.15, 0.20$ and $0.25$.
The plasma beta varies
from $\beta = 10^{-4}$ to $10^{-2}$ by controlling the electron thermal velocity.
Specifically, the plasma beta was set 
at $0.0001, 0.0003, 0.0010, 0.0030, 0.0100, 0.0200$ and $0.0300$.
Each beta value corresponds 
to $v_{te} / c = 0.007, 0.012, 0.022, 0.039, 0.071, 0.100$ and $0.122$, respectively, 
where $v_{te}$ is taken from the low density region and $c$ is speed of light.
Each numerical run (each data point
in subsequent Figs. (2)-(5) typically takes about 7-10
days on 64 parallel processors.
\section{Simulation Results}

\subsection{Source of the parallel electric field}

%

%

\begin{figure}
\includegraphics[scale = 0.5]{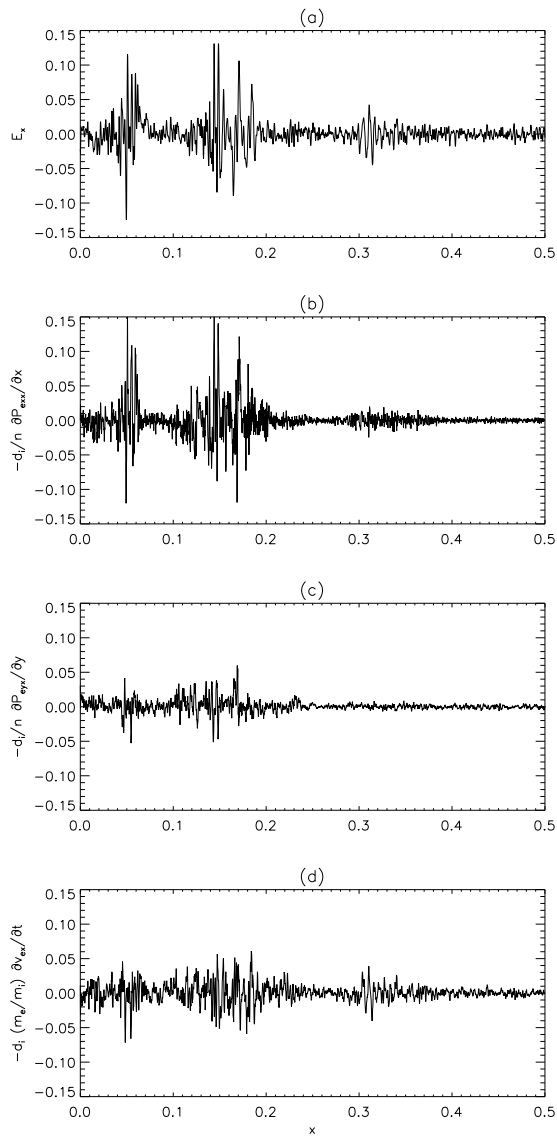}
  \caption
  {
Different term profiles along $x$-coordinate from the generalised Ohm's law:
    (a) $E_x$ (normalised to $v_{A0} B_0$),
    (b) $- (d_i / n) \partial P_{exx} / \partial x$, 
    (c) $- (d_i / n) \partial P_{eyx} / \partial y$, and 
    (d) $- d_i (m_e / m_i) \partial v_{ex} / \partial t$ 
    in the maximum density gradient region $y = 0.0103$ at $t = 0.4375$
    for the case, $\omega_d / \omega_{ci} = 0.3$, $\delta B / B_0 = 0.25$ and $\beta = 0.02$.
    The time derivative term in (d) is calculated from $t = 0.4370$ to $0.4375$
    (time interval corresponds to the inverse of electron plasma frequency $\omega^{-1}_{pe}$).
     Refer to text for the normalisation used.
  }
  \label{fig1}
\end{figure}

In order to understand details of the parallel electric field generation process, 
we now focus on the question: which term in the generalised Ohm's law is responsible for 
the generation of the parallel electric field?
The generalised Ohm's law can be written as
\begin{equation}
  \vec{E} = - \vec{v}_e \times \vec{B} 
            - \frac{\nabla \cdot \vec{P}_e}{n_e e}
            - \frac{m_e}{e} (\frac{\partial \vec{v}_e}{\partial t} 
            + \left( \vec{v}_e \cdot \nabla) \vec{v}_e \right), 
\label{g_ohms}
\end{equation}
where $\vec{E}$ and $\vec{B}$ are electric and magnetic fields, 
$\vec{v}$ is plasma velocity, $\vec{P}$ is pressure tensor ($3 \times 3$ matrix), 
$n$ is plasma number density, $m$ is  mass and $e$ is electric charge.
The subscript $e$ refers to an electron.
Normalising space coordinate by $L_x = 5000 \Delta$, 
fluid velocity by Alfv\'en speed (at the lowest density region)
$v_{A0}$, time by Alfv\'en transit time $\tau_A (= L / v_{A0})$, 
magnetic field by $B_0$, electric field by $v_{A0} B_0$,
number density by $n_0$ and pressure tensor by $B_0^2 / \mu_0$, 
a dimensionless version of Eq.~(\ref{g_ohms}) can be obtained
\begin{eqnarray}
  \vec{E} = - \vec{v}_e \times \vec{B} 
            - d_i \frac{\nabla \cdot \vec{P}_e}{n_e}
            - d_i \frac{m_e}{m_i} (\frac{\partial \vec{v}_e}{\partial t} 
            + \left( \vec{v}_e \cdot \nabla) \vec{v}_e \right), 
\label{nd_g_ohms}
\end{eqnarray}
where $d_i$ is the normalised ion skin depth ($d_i = c / \omega_{pi} L$).
Note that strictly speaking we should have used tildes in Eq.~(\ref{nd_g_ohms}) to denote 
dimensionless quantities, but we omit them for brevity.

Let us focus on the parallel electric field, $E_x$, 
which is generated  in the density gradient regions 
by phase mixing. It is given by, 
\begin{eqnarray}
\nonumber
  E_x = - (v_{ey} B_z - v_{ez} B_y) 
          - d_i \frac{1}{n} \left(
            \frac{\partial P_{exx}}{\partial x}
         + \frac{\partial P_{eyx}}{\partial y} \right) \\
          - d_i \frac{m_e}{m_i} \left(
              \frac{\partial v_{ex}}{\partial t}
            + v_{ex} \frac{\partial v_{ex}}{\partial x}
            + v_{ey} \frac{\partial v_{ex}}{\partial y} \right), 
\label{g_ohms_ex}
\end{eqnarray}
where $\partial / \partial z = 0$ is assumed because of spatially 2 dimensional model.

Fig.~\ref{fig1} shows the different term profiles (along the uniform
magnetic field in $x$-direction) in the generalised Ohm's law (Eq.~(\ref{g_ohms_ex})): 
(a) $E_x$,
(b) $- (d_i / n) \partial P_{exx} / \partial x$, 
(c) $- (d_i / n) \partial P_{eyx} / \partial y$, and 
(d) $- d_i (m_e / m_i) \partial v_{ex} / \partial t$ 
in the maximum density gradient region $y = 0.0103$ at $t = 0.4375$
for the case, $\omega_d / \omega_{ci} = 0.3$, $\delta B / B_0 = 0.25$ and $\beta = 0.02$.
The time derivative term in Fig.~\ref{fig1}(d) is calculated from $t = 0.4370$ to $0.4375$ 
(time interval corresponds to the inverse of electron plasma frequency $\omega^{-1}_{pe}$).
The other terms in right-hand side of Eq.~(\ref{g_ohms_ex}) are negligible.
In Fig.~\ref{fig1}(a), the parallel electric field is clearly observed 
in the density gradient regions where phase mixing can occur.
It should be noted that no parallel electric field is seen
away from the density gradient regions.
According to the generalised Ohm's law,
there has to be a term on the 
right-hand side of Eq.~(\ref{g_ohms_ex}) that supports this electric field.

By comparing Fig.~\ref{fig1}(a) to  Fig.~\ref{fig1}(b-d) it is clear that 
the parallel electric field (Fig.~\ref{fig1}(a)) is supported mostly by
the electron pressure tensor (Fig.~\ref{fig1}(b)), with a smaller contribution 
from the electron inertia term (Fig.~\ref{fig1}(d)). It is interesting to note that
previous results on collisionless reconnection both in tearing unstable 
Harris current sheet \cite{hesse99,birn01,pritchett01} and stressed X-point collapse 
\cite{th07,th08} have shown that the term in the generalised Ohm's law 
that is responsible for breaking the frozen-in condition, i.e. that
supports out-of-plane electric field is electron pressure tensor.
Here the electron pressure tensor supports (generates) the parallel
electric field. Thus, one should note a universal importance of
the electron pressure tensor in relation to supporting 
the electric fields in collisionless plasmas.

\subsection{Parametric study of the generated parallel electric 
field and the fraction of accelerated electrons}

To estimate how efficiently 
the parallel electric field is generated by phase mixing, 
as a function of model parameters, we introduce 
the average of the absolute value of the parallel electric field in the density gradient region:
\begin{equation}
  \frac{E^*}{E_0} = \frac{1}{L_x L_y} \int_{x = 0}^{L_x} 
  \int_{y = 26\Delta}^{77 \Delta} \frac{|E_x(x, y)|}{E_0} dx dy,
\label{e_star}
\end{equation}
where $E_0 = m_e c \omega_{pe}/ e$.
Note that in what follows the normalisation of the electric field is 
different from Section 3.1 where, usual for the
generalised Ohm's law, "fluid" normalisation ($v_{A0} B_0$) is used.
Normalisation $E_0 = m_e c \omega_{pe}/ e$ is usually referred to as "electrostatic".
By using the definition given by Eq.(\ref{e_star}) 
we can evaluate quantitatively the electric field generated 
in the density gradient region.
Although there are two density regions in our simulation box because of 
the periodic boundary condition, we focus on the lower one (the physics of the upper and 
the lower regions is essentially the same).
The range from $y = 26 \Delta$ to $77 \Delta$ indicates the 
density gradient. See \citet{2005A&A...435.1105T} for details.

Also, in order to investigate the fraction of accelerated electrons by the generated 
parallel electric field, we use particle data in the density gradient at lower side 
($26 \Delta \le y \le 77 \Delta$). We count 
the number of electrons with velocities greater than 
the thermal velocity ($v_{te}<v_x < c $) in the electron velocity 
distribution function, in the $x$-direction, 
at the final time snapshot $\omega_{ci} t = 54.69$,
and divide this by the total number of particles (with $0<v_x<c$)
in the same domain:
\begin{equation}
  \frac{N}{N_0} = \frac
    {
     \int_{v_x = v_{te}}^{c} \int_{x = 0}^{L_x} \int_{y = 26\Delta}^{77 \Delta} 
     f(v_x) dv_x dx dy
    }
    {
     \int_{v_x = 0}     ^{c} \int_{x = 0}^{L_x} \int_{y = 26\Delta}^{77 \Delta} 
     f(v_x) dv_x dx dy
    }.
\label{n_acc}
\end{equation}
Here it was to enough to integrate only positive region in this distribution
because electron acceleration was symmetrical in the $x$-direction.
Note that the initial velocity distribution function is nearly Maxwellian.
To maintain the balance of the total kinetic pressure throughout the system, 
the particle thermal velocity in the dense plasma region ($y = 100 \Delta$) is 
lower than the outside region ($y = 0 \Delta$ or $200 \Delta$) (see for details
Fig.(4) from \citet{2005A&A...435.1105T}).
However, in order to use Eq.~(\ref{n_acc}),
we had to estimate an appropriate velocity corresponding to
the thermal velocity in the Maxwellian.
Fortunately, initial electron velocity distribution function did not deviate much from 
the exact Maxwellian.
Therefore, we adopted the standard thermal velocity which was set to 
$36.8$\% of $f(v=0)$.
Recall that $f(v=v_{te}) = n_0 \exp(-v^2/v_{te}^2) = n_0 \exp(-1.0) = 0.368 n_0$, 
where $n_0$ is the peak number at $v = 0$ in Maxwellian.

\begin{figure}
\includegraphics[scale = 0.5]{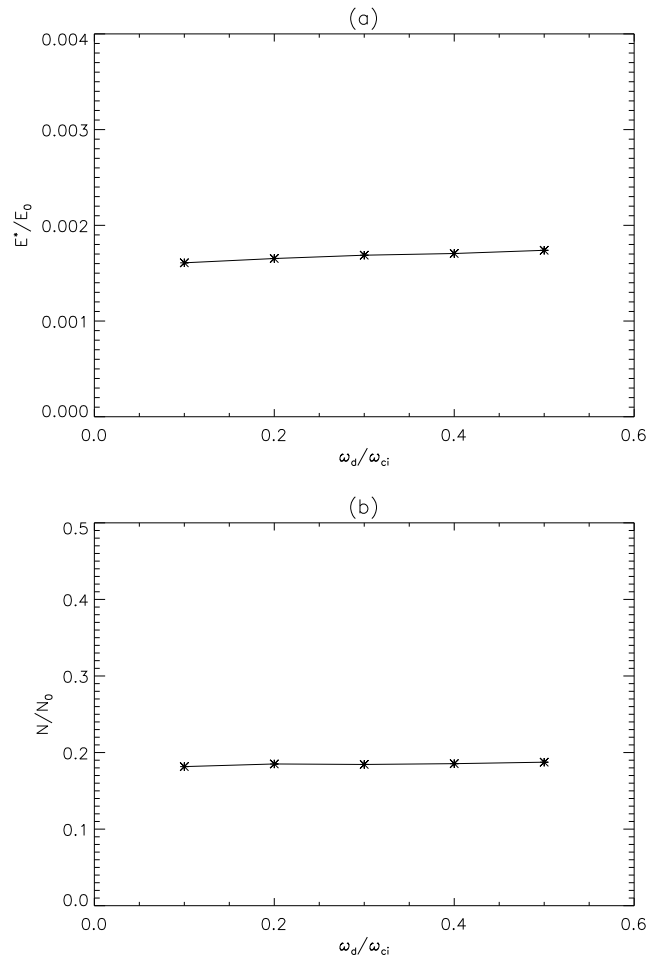}
  \caption{
           (a) $E^* / E_0$ vs. $\omega_d / \omega_{ci}$   and
           (b) $N / N_0$ vs. $\omega_d / \omega_{ci}$ 
           for $\delta B / B_0 = 0.05$ and $\beta = 0.02$.
          }
  \label{fig2}
\end{figure}

The choice of diagnostic for characterizing the degree of particle 
acceleration needs a further clarification.
Eq.(5) provides number of electrons with speed $v_x$ 
exceeding the thermal speed $v_{te}$ as a fraction of total distribution value.
However, in a Maxwellian plasma 16\% of the electrons have $v > v_{te}$ (for 
$v_{te} = \sqrt{2 k_B T_e/m_e}$). Thus, it may be tempting 
to either: (i) subtract this 16\% from our diagnostic in Eq.(5) or 
(ii) try to fit a Maxwellian to the electron distribution at the
final simulation time step (after all the acceleration has taken place)
and then count the number of electrons that have speeds $v_x > v_{te}$.
Our motivation to keep the definition given by Eq.(5) is two-fold:
(i) In context of the solar flare observations in X-rays, one always infers 
the integrated spectrum
averaged over some volume $V$, i.e. $<f(v)nV>$ (where $n$ is electron number density)  
and it is impossible to subtract the above mentioned 16\% without introducing 
additional uncertainties due to the unknown density and 
poorly known emitting volume, due to a line of sight effect (Dr. E. Kontar of 
University of Glasgow, private communication), see also Refs. \cite{bek03,bk05}.
Hence the definition given by Eq.(5) is more appropriate for comparison of
theory with the observations. Also, this is particularly timely
because the acceleration of electrons by Alfv\'enic waves in flares have been
recently studied \cite{2008ApJ...675.1645F}.
(ii) Although the electron distribution function at the initial time step is
nearly Maxwellian (despite density and temperature transverse inhomogeneities),
at the final stages of the simulation  the deviations from the 
Maxwellian form are large and hence a fit to a Maxwellian, in order to calculate
$v_{te}$, and in turn to count the super-thermal particles above that value ($v_x > v_{te}$),
is impractical.

Fig.~(\ref{fig2}) shows the generated parallel electric 
field, $E^*/E_0$, and the fraction of accelerated electrons,
$N/N_0$ as function of driving frequency of the IC wave (normalised to $\omega_{ci}$).
The wave frequency $\omega_d / \omega_{ci}$ was set at $0.1, 0.2, 0.3, 0.4$ 
and $0.5$. 
The wave amplitude and plasma beta were fixed at $\delta B / B_0 = 0.05$ and 
$\beta = 0.02$.
We gather from Fig.~(\ref{fig2}) that the generated parallel 
electric field and the fraction of accelerated electrons are 
independent of the IC wave
frequency remaining at a level of six orders of magnitude larger 
than the Dreicer value and approximately
20 \% respectively. This conclusion is based on the
following estimate  $E^*/E_0 \approx 0.0018$, i.e. $E^*=5473.4337$ V m$^{-1}$
(note that here solar coronal number density of 
 $n=10^{15}$ m$^{-3}$ was used). The Dreicer electric field (which
 is associated with the particle acceleration run-away regime \cite{dreicer}), 
$ E_d=(n e^3 \ln \Lambda)/(4 \pi \epsilon_0^2 k_B T)$, for $T=1$ MK (and hence 
 $\ln \Lambda=18.095$) is  $E_d=0.00547$ V m$^{-1}$. Thus, $E^*/E_d=10^6$.
 This result should be taken with caution because it is obtained
 with the ion-electron mass ratio of 16. As shown by \citet{2007NJPh....9..262T} (see their
 Figure 7) that attained amplitude of the generated parallel electric
 field scales strongly as  $\propto 1/(m_i/m_e)$.

\begin{figure}
\includegraphics[scale = 0.5]{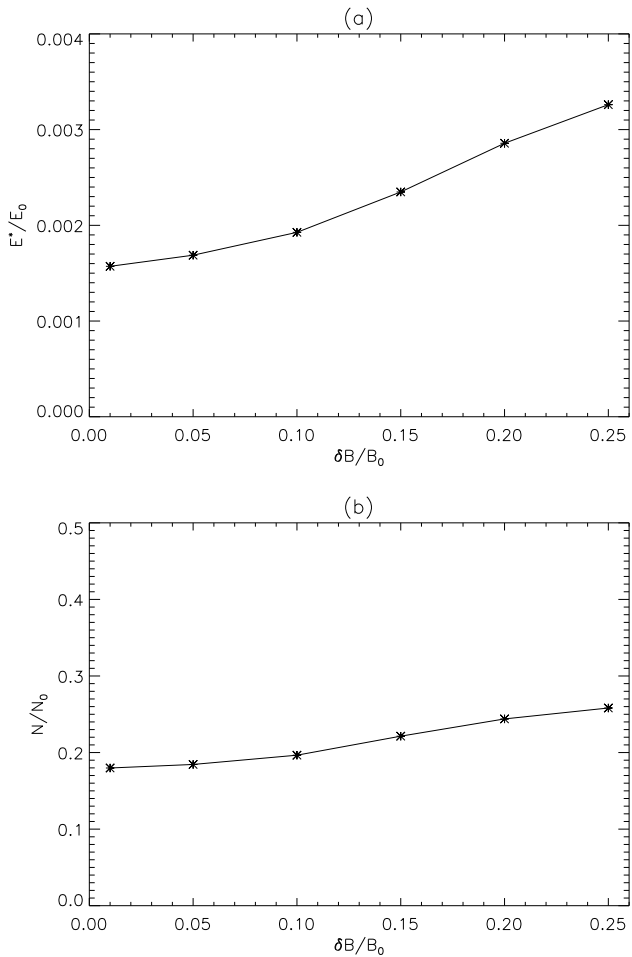}
  \caption{
           (a) $E^* / E_0$  vs. $\delta B / B_0$  and
           (b) $N / N_0$ vs.    $\delta B / B_0$  
           for $\omega_d / \omega_{ci} = 0.3$ and $\beta = 0.02$.
          }
  \label{fig3}
\end{figure}

Next, we explore how the generated parallel electric 
field and the fraction of accelerated electrons
depend on driving IC wave amplitude.
The latter, $\delta B / B_0$, was set at $0.01, 0.05, 0.10, 0.15, 0.20$ 
and $0.25$.
The wave frequency and plasma beta were fixed at $\omega_d / \omega_{ci} = 0.3$ 
and $\beta = 0.02$.
We gather from Fig.~\ref{fig3} that the generated parallel electric 
field and the fraction of accelerated electrons
 increase with the increase of IC wave  amplitude.
 This seems as a reasonable result because larger amplitude waves
 have more energy to give to electrons. Also, 
 non-linear effects would be progressively important.

\begin{figure}
\includegraphics[scale = 0.5]{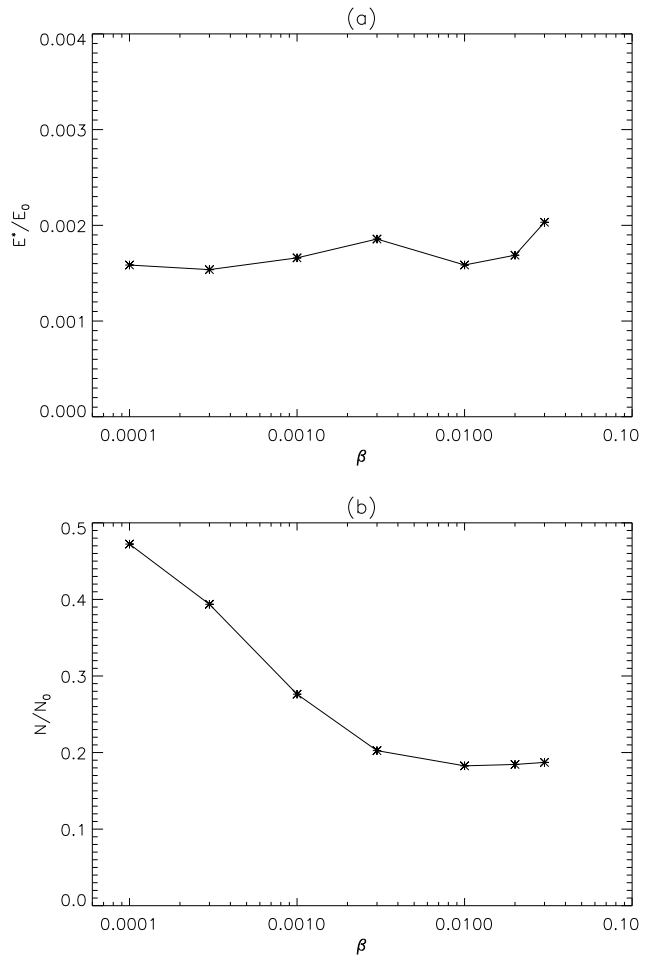}
  \caption{
           (a) $E^* / E_0$ vs. $\beta$   and
           (b) $N / N_0$ vs.  $\beta$  
               for $\omega_d / \omega_{ci} = 0.3$ and $\delta B / B_0 = 0.05$.
          }
  \label{fig4}
\end{figure}

We also explored the plasma beta dependence.
The plasma beta was set at $0.0001, 0.0003, 0.0010, 0.0030, 0.0100, 0.0200$ 
and $0.0300$.
The wave frequency and amplitude were fixed at $\omega_d / \omega_{ci} = 0.3$ 
and $\delta B / B_0 = 0.05$.
The plasma beta is defined as $\beta = 2 \mu_0 p / B^2 
= 2 (v_{te} / c)^2 / (\omega_{ce} / \omega_{pe})^2$.
We altered the plasma beta by changing the electron thermal velocity, 
affecting the plasma kinetic pressure.
Therefore, in this simulation, magnetic pressure is kept constant 
while plasma kinetic pressure varies.
Fig.~\ref{fig4}(a) does not show any correlation between plasma beta and 
the parallel electric field generation.
Incidentally, \citet{2002A&A...395..285T} investigated plasma beta dependence 
of the fast magnetosonic wave amplitude, which is
generated in a transversely inhomogeneous medium  when
an Alfv\'enic pulse is launched (using MHD numerical simulation).
According to Fig.~9(b) in  \citet{2002A&A...395..285T}, 
the maximum fast magnetosonic wave amplitude also does not depend on plasma beta.
\citet{2006A&A...455.1073T} alluded to the relation between
the non-linear fast magnetosonic wave and parallel electric field generation.
Hence it is not surprising that  in our Fig.~\ref{fig4}(a) we do not
see plasma beta dependence. What is surprising, at the first glance,
in Fig.~\ref{fig4}(b), is that the fraction of accelerated particles strongly
depends on beta. In particular, a decrease in beta (for $\beta = 0.0001$) yields as
large percentage as  $N / N_0 = 0.472 \sim 47$\%!
One can conjecture that this is due to the fact the in the
case of small plasma beta, magnetic effects dominate
over thermal ones, and because IC wave is essentially a magnetic-type
perturbation, electrons respond better to the wave influence and
accelerate more efficiently.

\subsection{Amplitude decay law in the kinetic regime}

\begin{figure}
\includegraphics[scale = 0.5]{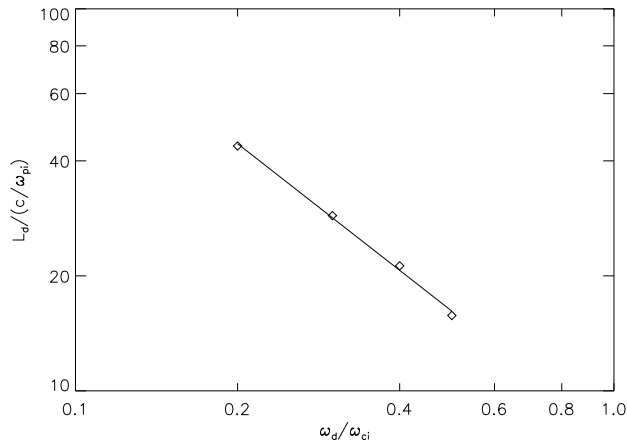}
  \caption{
           A log-log plot of the
	   dissipation length $L_d / (c/\omega_{pi})$
	    vs. driving IC wave frequency $\omega_d / \omega_{ci}$ (open symbols). 
           The solid line is the least squares fit with a slope of $-1.10$.
                     }
  \label{fig5}
\end{figure}

\citet{2005A&A...435.1105T} established that in the collisionless, kinetic regime
Alfv\'en (IC) wave amplitude in density gradient region decays with
distance (from where it is driven) according to collisional MHD formula of 
\citet{1983A&A...117..220H}
\begin{equation}
  B_z \sim \exp \left[- 
  \frac{\eta \omega_d^2 (\partial_y v_A)^2} {6 v_A^5} x^3 \right],
  \label{mhd_decay_law}
\end{equation}
where $\eta$ is resistivity (divided by $\mu_0$, i.e. by $\eta$
we mean $1/(\sigma \mu_0)$, $\omega_d$ is (driving) wave frequency, $v_A$ is 
Alfv\'en velocity and $x$ is the axis which AW propagates.
Recall that plasma density inhomogeneity is a function of $y$ and 
wave propagates in the $x$ direction.
In particular, (see for details left panel of Fig.6 in \citet{2005A&A...435.1105T})
it was shown that, at a fixed time instance corresponding to well-developed phase mixing,
$B_z(x) \propto \exp\left(-(x/L_d)^3\right)$. In other words, one can define
an empirical dissipation length, $L_d$, according to collisional MHD formula 
(Eq.(\ref{mhd_decay_law})) in the
collisionless, kinetic regime. Applicability of the MHD formula
in the kinetic regime was rather surprising.
We now turn back to the previous results in order to see how far this
MHD-kinetic analogy can be stretched. In particular,
one can e.g. check the $\omega_d^2$ scaling in the Eq.(\ref{mhd_decay_law}).
For this purpose, after simple algebra applied to $B_z \propto \exp[-(x/L_d)^3]$, 
for fixed $\eta, \partial_y v_A$ and $v_A$,
one can obtain the following scaling $\log_{10} L_d = -2/3 \log_{10} \omega_d + C$,
where $C$ is some constant.
Hence the slope of a log-log graph of
dissipation length $L_d$ versus $\omega_d$ is expected to be $-2/3 = -0.67$.
Fig.~\ref{fig5} presents the dissipation length (distance over which the wave damps)
dependence on the IC wave frequency.
The values in Fig.~\ref{fig5} were obtained from line data of magnetic field 
$B_z$ in the maximum plasma gradient $Y / (c / \omega_{pe}) = 14.8$ at 
$\omega_{ci} t = 82.0$ for $\omega_d / \omega_{ci} = 0.2$, 
$\omega_{ci} t = 54.7$ for $\omega_d / \omega_{ci} = 0.3$, 
$\omega_{ci} t = 41.0$ for $\omega_d / \omega_{ci} = 0.4$ and 
$\omega_{ci} t = 32.8$ for $\omega_d / \omega_{ci} = 0.5$, respectively.
The reason for the different snapshot times
is that we only consider well-developed phase mixing, i.e. when IC wave
is fully damped. As in left panel of Fig.6 in \citet{2005A&A...435.1105T})
we fit $B_z(x)$ to $\propto \exp\left(-(x/L_d)^3\right)$ and obtain empirical dissipation length $L_d$.
 we gather from Fig.~\ref{fig5} that the slope is $-1.10$ contrary to the above
 prediction of $-2/3 = -0.67$.
 
 To address the inconsistency we conjecture that the resistivity might be variable.
One can estimate the resistivity for each case of driving wave frequency considered,
by calculating $\eta = 6 v_A^5 / (\omega_d^2 (\partial_y v_A)^2 L_d^3)$.
We reiterate that strictly speaking PIC
simulation code is collisionless and hence no resistive effects
exist. However, scattering of particles by magnetic
fields plays effective role of collisions. When resistivity is mentioned
we refer to "effective" resistivity.
Normalising the frequency by IC frequency $\omega_{ci}$,
length by ion skin depth $c / \omega_{pi}$ and 
velocity by speed of light $c$,
the dimensionless resistivity is given by, 
\begin{eqnarray}
  \frac{\eta}{c^2 \omega_{pi}/ \omega_{ci}^2} &=&
  \frac{2}{27} 
  \left( \frac{\omega_d}{\omega_{ci}} \right)^{-2}
  \left( \frac{L}{c / \omega_{pi}}    \right)^2
  \left( \frac{L_d}{c / \omega_{pi}}  \right)^{-3}
  \left( \frac{v_A}{c}                \right)^3 \nonumber \\
  &&
  \sqrt{1 + 3 \exp \left[- \left(
  \frac{y - y_c}{L}\right)^6 \right]} \nonumber \\
  &&
  \exp \left[2 \left( \frac{y - y_c}{L} \right)^6 \right]
  \left( \frac{y - y_c}{L} \right)^{-10}.
  \label{eta}
\end{eqnarray}
We can now put in the known parameters $L / (c / \omega_{pi}) = 1.25$, 
$v_A / c = 0.25$ and $(y - y_c) / L = (148.5 - 100) / 50 = 0.97$ 
into Eq.~(\ref{eta}).
Here $L$ and $y_c$ are the width and the centre of plasma density gradient, 
respectively.
Therefore the dimensionless resistivity can be estimated using
\begin{equation}
  \frac{\eta}{c^2 \omega_{pi}/ \omega_{ci}^2} = 1.97 \times 10^{-2}
  \left( \frac{\omega_d}{\omega_{ci}} \right)^{-2}
  \left( \frac{L_d}{c / \omega_{pi}}  \right)^{-3}.
  \label{eta_pic}
\end{equation}
As the driving wave frequency and the dissipation length for each case are given by choice and empirically,
respectively, one can obtain the resistivity by substituting above values into 
Eq.~(\ref{eta_pic}).
The result is $\eta / (c^2 \omega_{pi}/ \omega_{ci}^2) = 5.88 \times 10^{-6}$ 
(for $\omega_d / \omega_{ci} = 0.2$), 
$9.21 \times 10^{-6}$ (for $\omega_d / \omega_{ci}=0.3$), 
$1.28 \times 10^{-5}$ (for $\omega_d / \omega_{ci}=0.4$) and 
$2.02 \times 10^{-5}$ (for $\omega_d / \omega_{ci}=0.5$).
Thus our initial conjecture that the effective resistivity depends on the
driving IC wave frequency turns out to be correct.
But the main conclusion of this analysis is that 
despite collisional MHD scaling Eq.(\ref{mhd_decay_law}) being applicable
to the collisionless, kinetic regime of phase mixing, i.e. 
$B_z(x) \propto \exp\left(-(x/L_d)^3\right)$ scaling holds,
stretching the MHD-kinetic analogy further to $\omega_d^2$ dependence
under the exponent
is not valid (due to the effective resistivity being a function of $\omega_d$).

\subsection{Effective  anomalous resistivity}

Issue of anomalous resistivity is central for many space and laboratory 
plasma applications. It can facilitate fast magnetic reconnection via Petschek type
mechanism (if $\eta$ is not spatially uniform), or have significant implications for wave heating models of
solar corona where normal Spitzer resistivity is too small to produce any sizable effect.
Ref.\cite{anres} presented plasma resistivity measurements in the reconnection 
current sheet of Magnetic Reconnection 
Experiment (MRX) (see ref.\cite{yamada97} for details of the
experimental setup). They established that in the collisionless regime
measured resistivity values can be more than an order of magnitude larger than the
Spitzer value \cite{anres}.

Let us apply our PIC simulation results to see
if there is any evidence for the anomalous effective
resistivity. We fix physical parameters corresponding to 
solar coronal plasmas: $B = 0.01 T$ and 
plasma number density $n_0 = 2 \times 10^{15} m^{-3}$, i.e.
$\omega_{ci} = eB / (16 m_e) = 1.10 \times 10^8 rad/s$ and 
$\omega_{pi} = \sqrt{n_0 e^2 / (16 m_e \epsilon_0)} = 6.30 \times 10^8 rad/s$.
Eq.~(\ref{eta_pic}) can be rewritten as:
\begin{equation}
  \eta = 9.24 \times 10^{7}
  \left( \frac{\omega_d}{\omega_{ci}} \right)^{-2}
  \left( \frac{L_d}{c / \omega_{pi}}  \right)^{-3}.
  \label{eta_real}
\end{equation}
Similarly to the previous calculation in Eq.~(\ref{eta_pic}), 
now using Eq.~(\ref{eta_real}) we  obtain
$\eta = 2.76 \times 10^4$ (for $\omega_d/ \omega_{ci} = 0.2$),
$4.32 \times 10^4$ (for $\omega_d/ \omega_{ci} =0.3)$,
$6.02 \times 10^4$ (for $\omega_d/ \omega_{ci} =0.4)$ and
$9.46 \times 10^4$ (for $\omega_d/ \omega_{ci} =0.5)$.
Here units of the resistivity are  m$^2$ sec$^{-1}$.
We gather that all values are in the range of $\approx 10^4-10^5$ m$^2$ sec$^{-1}$.
Spitzer resistivity (normalised to 
$\mu_0$) for the above parameters and $T=1$ MK is 1.83 m$^2$ sec$^{-1}$.
Thus, we conclude that our numerical simulations provide
effective resistivity values of $\approx 10^4$ times
larger than Spitzer value, which is indicative of
the anomalous resistivity.
It should be mentioned that these results were obtained for
the ion-to-electron mass ratio of 16. Clearly one
would expect some dependence of the obtained effective resistivity 
on the mass ratio. Thus, the obtained results should be taken with caution.
\section{Conclusions}

Let us summarise the above findings:

We used the generalised Ohm's law and found that the parallel electric field, 
which is generated by propagation of IC (Alfv\'enic) wave in a transversely inhomogeneous plasma,
is supported mostly by
the electron pressure tensor, with a smaller contribution from the electron inertia term.
Surprisingly, this result resembles closely to the 
previous results on collisionless reconnection both in tearing unstable 
Harris current sheet \cite{hesse99,birn01,pritchett01} and stressed X-point collapse 
\cite{th07,th08}. However, in the latter two cases, the generated electric field is in
the plane perpendicular to the magnetic field.
Thus, a universal importance of
the  electron pressure tensor in relation to supporting 
the electric fields in collisionless plasmas should be noted.

We explored physical parameter space of the problem with regards
to the efficiency of generation of parallel electric field
and acceleration of electrons. We found that
the generated parallel electric field and the fraction of accelerated electrons are 
independent of the IC wave
 frequency staying at a level that is $10^6$ times larger 
 than the Dreicer value and approximately
20 \% respectively. The generated parallel electric field and the fraction of accelerated electrons 
increase with the increase of IC wave  amplitude. The generated parallel electric field seems to be
independent of  plasma-$\beta$. However, the fraction of accelerated electrons strongly increases 
with the decrease of plasma-$\beta$, e.g.  for plasma $\beta=0.0001$ the fraction of accelerated electrons
can be as large as 47 \%.

Previously it was established  that in the collisionless, kinetic regime
phase-mixed Alfv\'en (IC) wave amplitude damps with distance of propagation
according to $\propto \exp[-(x/L_d)^3]$ \cite{2005A&A...435.1105T}, which
resembles closely to collisional MHD result of \citet{1983A&A...117..220H}.
We tried to stretch this analogy further by investigating 
how the dissipation length $L_d$ scales with the IC driving frequency.
We found that the scaling is different from the MHD result.
We have shown that this discrepancy can be attributed to 
the frequency dependence of the effective resistivity.

We also found that  the effective resistivity, albeit for unrealistic mass ratio,
still is as large as $10^4$ times the classical Spitzer value.

\begin{acknowledgements}
Numerical calculations of this work were performed using the MHD Cluster 
at University of St Andrews. 
Author acknowledges useful discussion of solar flare 
observational aspects with Dr. E. Kontar.
This research was supported by the United Kingdom's 
Science and Technology Facilities Council (STFC).
\end{acknowledgements}


\begin{thebibliography}{28}
\expandafter\ifx\csname natexlab\endcsname\relax\def\natexlab#1{#1}\fi
\expandafter\ifx\csname bibnamefont\endcsname\relax
  \def\bibnamefont#1{#1}\fi
\expandafter\ifx\csname bibfnamefont\endcsname\relax
  \def\bibfnamefont#1{#1}\fi
\expandafter\ifx\csname citenamefont\endcsname\relax
  \def\citenamefont#1{#1}\fi
\expandafter\ifx\csname url\endcsname\relax
  \def\url#1{\texttt{#1}}\fi
\expandafter\ifx\csname urlprefix\endcsname\relax\def\urlprefix{URL }\fi
\providecommand{\bibinfo}[2]{#2}
\providecommand{\eprint}[2][]{\url{#2}}

\bibitem[{\citenamefont{{Uberoi}}(1972)}]{1972PhFl...15.1673U}
\bibinfo{author}{\bibfnamefont{C.}~\bibnamefont{{Uberoi}}},
  \bibinfo{journal}{Phys. Fluids} \textbf{\bibinfo{volume}{15}},
  \bibinfo{pages}{1673} (\bibinfo{year}{1972}).

\bibitem[{\citenamefont{{Tataronis} and {Grossmann}}(1973)}]{1973ZPhy..261..203T}
\bibinfo{author}{\bibfnamefont{J.}~\bibnamefont{{Tataronis}}} \bibnamefont{and}
  \bibinfo{author}{\bibfnamefont{W.}~\bibnamefont{{Grossmann}}},
  \bibinfo{journal}{Zeitschrift fur Physik} \textbf{\bibinfo{volume}{261}},
  \bibinfo{pages}{203} (\bibinfo{year}{1973}).

\bibitem[{\citenamefont{{Grossmann} and {Tataronis}}(1973)}]{1973ZPhy..261..217G}
\bibinfo{author}{\bibfnamefont{W.}~\bibnamefont{{Grossmann}}} \bibnamefont{and}
  \bibinfo{author}{\bibfnamefont{J.}~\bibnamefont{{Tataronis}}},
  \bibinfo{journal}{Zeitschrift fur Physik} \textbf{\bibinfo{volume}{261}},
  \bibinfo{pages}{217} (\bibinfo{year}{1973}).

\bibitem[{\citenamefont{{Hasegawa} and {Chen}}(1974)}]{1974PhRvL..32..454H}
\bibinfo{author}{\bibfnamefont{A.}~\bibnamefont{{Hasegawa}}} \bibnamefont{and}
  \bibinfo{author}{\bibfnamefont{L.}~\bibnamefont{{Chen}}},
  \bibinfo{journal}{Phys. Rev. Lett.} \textbf{\bibinfo{volume}{32}},
  \bibinfo{pages}{454} (\bibinfo{year}{1974}).

\bibitem[{\citenamefont{{Chen} and {Hasegawa}}(1974)}]{1974PhFl...17.1399C}
\bibinfo{author}{\bibfnamefont{L.}~\bibnamefont{{Chen}}} \bibnamefont{and}
  \bibinfo{author}{\bibfnamefont{A.}~\bibnamefont{{Hasegawa}}},
  \bibinfo{journal}{Phys. Fluids} \textbf{\bibinfo{volume}{17}},
  \bibinfo{pages}{1399} (\bibinfo{year}{1974}).

\bibitem[{\citenamefont{{Tataronis}}(1975)}]{1975JPlPh..13...87T}
\bibinfo{author}{\bibfnamefont{J.~A.} \bibnamefont{{Tataronis}}},
  \bibinfo{journal}{J. Plasma Phys.} \textbf{\bibinfo{volume}{13}},
  \bibinfo{pages}{87} (\bibinfo{year}{1975}).

\bibitem[{\citenamefont{{Heyvaerts} and {Priest}}(1983)}]{1983A&A...117..220H}
\bibinfo{author}{\bibfnamefont{J.}~\bibnamefont{{Heyvaerts}}} \bibnamefont{and}
  \bibinfo{author}{\bibfnamefont{E.~R.} \bibnamefont{{Priest}}},
  \bibinfo{journal}{Astron. Astrophys.} \textbf{\bibinfo{volume}{117}},
  \bibinfo{pages}{220} (\bibinfo{year}{1983}).

\bibitem[{\citenamefont{{G\'enot} et~al.}(1999)\citenamefont{{G\'enot},  {Louarn}, and {Le Qu\'eau}}}]{1999JGR...10422649G}
\bibinfo{author}{\bibfnamefont{V.}~\bibnamefont{{G\'enot}}},
  \bibinfo{author}{\bibfnamefont{P.}~\bibnamefont{{Louarn}}}, \bibnamefont{and}
  \bibinfo{author}{\bibfnamefont{D.}~\bibnamefont{{Le Qu\'eau}}},
  \bibinfo{journal}{J. \ Geophys. \ Res.} \textbf{\bibinfo{volume}{104}},
  \bibinfo{pages}{22649} (\bibinfo{year}{1999}).

\bibitem[{\citenamefont{{G\'enot} et~al.}(2004)\citenamefont{{G\'enot},   {Louarn}, and {Mottez}}}]{2004AnGeo..22.2081G}
\bibinfo{author}{\bibfnamefont{V.}~\bibnamefont{{G\'enot}}},
  \bibinfo{author}{\bibfnamefont{P.}~\bibnamefont{{Louarn}}}, \bibnamefont{and}
  \bibinfo{author}{\bibfnamefont{F.}~\bibnamefont{{Mottez}}},
  \bibinfo{journal}{Ann. Geophysicae} \textbf{\bibinfo{volume}{22}},
  \bibinfo{pages}{2081} (\bibinfo{year}{2004}).

\bibitem[{\citenamefont{{Tsiklauri}   et~al.}(2005{\natexlab{a}})\citenamefont{{Tsiklauri}, {Sakai}, and {Saito}}}]{2005A&A...435.1105T}
\bibinfo{author}{\bibfnamefont{D.}~\bibnamefont{{Tsiklauri}}},
  \bibinfo{author}{\bibfnamefont{J.-I.} \bibnamefont{{Sakai}}},
  \bibnamefont{and} \bibinfo{author}{\bibfnamefont{S.}~\bibnamefont{{Saito}}},
  \bibinfo{journal}{Astron. Astrophys.} \textbf{\bibinfo{volume}{435}},
  \bibinfo{pages}{1105} (\bibinfo{year}{2005}{\natexlab{a}}).

\bibitem[{\citenamefont{{Tsiklauri} et~al.}(2005{\natexlab{b}})\citenamefont{{Tsiklauri}, {Sakai}, and {Saito}}}]{2005NJPh....7...79T}
\bibinfo{author}{\bibfnamefont{D.}~\bibnamefont{{Tsiklauri}}},
  \bibinfo{author}{\bibfnamefont{J.-I.} \bibnamefont{{Sakai}}},
  \bibnamefont{and} \bibinfo{author}{\bibfnamefont{S.}~\bibnamefont{{Saito}}},
  \bibinfo{journal}{New J. Physics} \textbf{\bibinfo{volume}{7}},
  \bibinfo{pages}{79} (\bibinfo{year}{2005}{\natexlab{b}}).

\bibitem[{\citenamefont{{Mottez} et~al.}(2006)\citenamefont{{Mottez}, {G{\'e}not}, and {Louarn}}}]{2006A&A...449..449M}
\bibinfo{author}{\bibfnamefont{F.}~\bibnamefont{{Mottez}}},
  \bibinfo{author}{\bibfnamefont{V.}~\bibnamefont{{G{\'e}not}}},
  \bibnamefont{and} \bibinfo{author}{\bibfnamefont{P.}~\bibnamefont{{Louarn}}},
  \bibinfo{journal}{Astron. Astrophys.} \textbf{\bibinfo{volume}{449}},
  \bibinfo{pages}{449} (\bibinfo{year}{2006}).

\bibitem[{\citenamefont{{Tsiklauri}}(2007)}]{2007NJPh....9..262T}
\bibinfo{author}{\bibfnamefont{D.}~\bibnamefont{{Tsiklauri}}},
  \bibinfo{journal}{New J. Physics} \textbf{\bibinfo{volume}{9}},
  \bibinfo{pages}{262} (\bibinfo{year}{2007}).

\bibitem[{\citenamefont{{Tsiklauri}}(2006)}]{2006A&A...455.1073T}
\bibinfo{author}{\bibfnamefont{D.}~\bibnamefont{{Tsiklauri}}},
  \bibinfo{journal}{Astron. Astrophys.} \textbf{\bibinfo{volume}{455}},
  \bibinfo{pages}{1073} (\bibinfo{year}{2006}).

\bibitem[{\citenamefont{{Pritchett}}(2001)}]{pritchett01}
\bibinfo{author}{\bibfnamefont{P.~L.} \bibnamefont{{Pritchett}}},
  \bibinfo{journal}{J. \ Geophys. \ Res.} \textbf{\bibinfo{volume}{106}},
  \bibinfo{pages}{3783} (\bibinfo{year}{2001}).

\bibitem[{\citenamefont{{Tsiklauri} and {Haruki}}(2008)}]{th08}
\bibinfo{author}{\bibfnamefont{D.}~\bibnamefont{{Tsiklauri}}} \bibnamefont{and}
  \bibinfo{author}{\bibfnamefont{T.}~\bibnamefont{{Haruki}}},
  \bibinfo{journal}{{\it "Physics of collisionless reconnection in a stressed
  X-point collapse"}, Physics of Plasmas (in press)}  (\bibinfo{year}{2008}).

\bibitem[{\citenamefont{{Fletcher}}(2005)}]{2005SSRv..121..141F}
\bibinfo{author}{\bibfnamefont{L.}~\bibnamefont{{Fletcher}}},
  \bibinfo{journal}{Space Science Reviews} \textbf{\bibinfo{volume}{121}},
  \bibinfo{pages}{141} (\bibinfo{year}{2005}).

\bibitem[{\citenamefont{Moran}(2001)}]{moran01}
\bibinfo{author}{\bibfnamefont{T.~G.} \bibnamefont{Moran}},
  \bibinfo{journal}{Astron. \ Astrophys.} \textbf{\bibinfo{volume}{374}},
  \bibinfo{pages}{L9} (\bibinfo{year}{2001}).

\bibitem[{\citenamefont{{Hesse} et~al.}(1999)\citenamefont{{Hesse}, {Schindler}, {Birn}, and {Kuznetsova}}}]{hesse99}
\bibinfo{author}{\bibfnamefont{M.}~\bibnamefont{{Hesse}}},
  \bibinfo{author}{\bibfnamefont{K.}~\bibnamefont{{Schindler}}},
  \bibinfo{author}{\bibfnamefont{J.}~\bibnamefont{{Birn}}}, \bibnamefont{and}
  \bibinfo{author}{\bibfnamefont{M.}~\bibnamefont{{Kuznetsova}}},
  \bibinfo{journal}{Phys. Plasmas} \textbf{\bibinfo{volume}{6}},
  \bibinfo{pages}{1781} (\bibinfo{year}{1999}).

\bibitem[{\citenamefont{{Birn} et~al.}(2001)\citenamefont{{Birn}, {Drake}, {Shay}, {Rogers}, {Denton}, {Hesse}, {Kuznetsova}, {Ma},{Bhattacharjee}, {Otto} }}]{birn01}
\bibinfo{author}{\bibfnamefont{J.}~\bibnamefont{{Birn}}},
  \bibinfo{author}{\bibfnamefont{J.~F.} \bibnamefont{{Drake}}},
  \bibinfo{author}{\bibfnamefont{M.~A.} \bibnamefont{{Shay}}},
  \bibinfo{author}{\bibfnamefont{B.~N.} \bibnamefont{{Rogers}}},
  \bibinfo{author}{\bibfnamefont{R.~E.} \bibnamefont{{Denton}}},
  \bibinfo{author}{\bibfnamefont{M.}~\bibnamefont{{Hesse}}},
  \bibinfo{author}{\bibfnamefont{M.}~\bibnamefont{{Kuznetsova}}},
  \bibinfo{author}{\bibfnamefont{Z.~W.} \bibnamefont{{Ma}}},
  \bibinfo{author}{\bibfnamefont{A.}~\bibnamefont{{Bhattacharjee}}},
  \bibinfo{author}{\bibfnamefont{A.}~\bibnamefont{{Otto}}}, 
  \bibinfo{journal}{J. \ Geophys. \ Res.}
  \textbf{\bibinfo{volume}{106}}, \bibinfo{pages}{3715} (\bibinfo{year}{2001}).

\bibitem[{\citenamefont{{Tsiklauri} and {Haruki}}(2007)}]{th07}
\bibinfo{author}{\bibfnamefont{D.}~\bibnamefont{{Tsiklauri}}} \bibnamefont{and}
  \bibinfo{author}{\bibfnamefont{T.}~\bibnamefont{{Haruki}}},
  \bibinfo{journal}{Phys. Plasmas} \textbf{\bibinfo{volume}{14}},
  \bibinfo{pages}{112905} (\bibinfo{year}{2007}).

\bibitem[{\citenamefont{{Brown} et~al.}(2003)\citenamefont{{Brown}, {Emslie},  and {Kontar}}}]{bek03}
\bibinfo{author}{\bibfnamefont{J.~C.} \bibnamefont{{Brown}}},
  \bibinfo{author}{\bibfnamefont{A.~G.} \bibnamefont{{Emslie}}},
  \bibnamefont{and} \bibinfo{author}{\bibfnamefont{E.~P.}
  \bibnamefont{{Kontar}}}, \bibinfo{journal}{Astrophys. J.}
  \textbf{\bibinfo{volume}{595}}, \bibinfo{pages}{L115} (\bibinfo{year}{2003}).

\bibitem[{\citenamefont{{Brown} and {Kontar}}(2005)}]{bk05}
\bibinfo{author}{\bibfnamefont{J.~C.} \bibnamefont{{Brown}}} \bibnamefont{and}
  \bibinfo{author}{\bibfnamefont{E.~P.} \bibnamefont{{Kontar}}},
  \bibinfo{journal}{Adv. Space Res.} \textbf{\bibinfo{volume}{35}},
  \bibinfo{pages}{1675} (\bibinfo{year}{2005}).

\bibitem[{\citenamefont{{Fletcher} and {Hudson}}(2008)}]{2008ApJ...675.1645F}
\bibinfo{author}{\bibfnamefont{L.}~\bibnamefont{{Fletcher}}} \bibnamefont{and}
  \bibinfo{author}{\bibfnamefont{H.~S.} \bibnamefont{{Hudson}}},
  \bibinfo{journal}{Astrophys. J.} \textbf{\bibinfo{volume}{675}},
  \bibinfo{pages}{1645} (\bibinfo{year}{2008}).

\bibitem[{\citenamefont{Dreicer}(1959)}]{dreicer}
\bibinfo{author}{\bibfnamefont{H.}~\bibnamefont{Dreicer}},
  \bibinfo{journal}{Phys. Rev.} \textbf{\bibinfo{volume}{115}},
  \bibinfo{pages}{238} (\bibinfo{year}{1959}).

\bibitem[{\citenamefont{{Tsiklauri} et~al.}(2002)\citenamefont{{Tsiklauri},   {Nakariakov}, and {Arber}}}]{2002A&A...395..285T}
\bibinfo{author}{\bibfnamefont{D.}~\bibnamefont{{Tsiklauri}}},
  \bibinfo{author}{\bibfnamefont{V.~M.} \bibnamefont{{Nakariakov}}},
  \bibnamefont{and} \bibinfo{author}{\bibfnamefont{T.~D.}
  \bibnamefont{{Arber}}}, \bibinfo{journal}{Astron. Astrophys.}
  \textbf{\bibinfo{volume}{395}}, \bibinfo{pages}{285} (\bibinfo{year}{2002}).

\bibitem[{\citenamefont{{Trintchouk} et~al.}(2003)\citenamefont{{Trintchouk}, {Yamada}, {Ji}, {Kulsrud}, and {Carter}}}]{anres}
\bibinfo{author}{\bibfnamefont{F.}~\bibnamefont{{Trintchouk}}},
  \bibinfo{author}{\bibfnamefont{M.}~\bibnamefont{{Yamada}}},
  \bibinfo{author}{\bibfnamefont{H.}~\bibnamefont{{Ji}}},
  \bibinfo{author}{\bibfnamefont{R.~M.} \bibnamefont{{Kulsrud}}},
  \bibnamefont{and} \bibinfo{author}{\bibfnamefont{T.~A.}
  \bibnamefont{{Carter}}}, \bibinfo{journal}{Phys. Plasmas}
  \textbf{\bibinfo{volume}{10}}, \bibinfo{pages}{319} (\bibinfo{year}{2003}).

\bibitem[{\citenamefont{{Yamada} et~al.}(1997)\citenamefont{{Yamada}, {Ji}, {Hsu}, {Carter}, {Kulsrud}, {Bretz}, {Jobes}, {Ono}, and {Perkins}}}]{yamada97} 
\bibinfo{author}{\bibfnamefont{M.}~\bibnamefont{{Yamada}}},
  \bibinfo{author}{\bibfnamefont{H.}~\bibnamefont{{Ji}}},
  \bibinfo{author}{\bibfnamefont{S.}~\bibnamefont{{Hsu}}},
  \bibinfo{author}{\bibfnamefont{T.}~\bibnamefont{{Carter}}},
  \bibinfo{author}{\bibfnamefont{R.}~\bibnamefont{{Kulsrud}}},
  \bibinfo{author}{\bibfnamefont{N.}~\bibnamefont{{Bretz}}},
  \bibinfo{author}{\bibfnamefont{F.}~\bibnamefont{{Jobes}}},
  \bibinfo{author}{\bibfnamefont{Y.}~\bibnamefont{{Ono}}}, \bibnamefont{and}
  \bibinfo{author}{\bibfnamefont{F.}~\bibnamefont{{Perkins}}},
  \bibinfo{journal}{Phys. Plasmas} \textbf{\bibinfo{volume}{4}},
  \bibinfo{pages}{1936} (\bibinfo{year}{1997}).

\end{thebibliography}

\end{document}